\documentclass[epj,twocolumn]{webofc} 
\usepackage[varg]{txfonts}   
\usepackage{lineno}
\woctitle{ISVHECRI 2018}
\begin{document}
\title{Detection of Cosmic-Ray Ensembles with CREDO}
%
%

\author{\firstname{Krzysztof W. Wo\'{z}niak\inst{1}} 
	\firstname{for the CREDO Collaboration}\inst{2}\fnsep\thanks{The CREDO Collaboration: K. Almeida-Cheminant,
\L{}.~Bratek,
D.A.~Castillo,
N.~Dhital,
A.R.~Duffy,
D.~G\'{o}ra,
B.~Hnatyk,
P.~Homola,
P.~Jagoda,
J.~Ja\l{}ocha-Bratek,
M.~Kasztelan,
D.~Lema\'{n}ski,
P.~Kov\'{a}cs,
V.~Nazari,
M.~Nied\'{z}wiecki,
K.~Smelcerz,
K.~Smolek,
J.~Stasielak,
S.~Stuglik,
O.~Sushchov,
K.W.~Wo\'{z}niak,
J.~Zamora-Sa\'{a}.
}}
	

\institute{	Institute of Nuclear Physics, Polish Acedemy of Sciences
	Krakow, Poland\\
	\email{krzysztof.wozniak@ifj.edu.pl}	
\and 
website: http://credo.science
          }

\abstract{%
  One of the main objectives of cosmic-ray studies are precise measurements of energy and chemical composition of particles with extreme  energies. Large and sophisticated detectors are used to find events seen as showers starting in the Earth's atmosphere with recorded energies larger than 100 EeV. However, a Cosmic-Ray Ensemble (CRE) developing before reaching the Earth as a bunch of correlated particles may spread over larger areas and requires an extended set of detectors to be discovered. The Cosmic-Ray Extremely Distributed Observatory (CREDO) is a solution to find such phenomena.  Even simple detectors measuring the particle arrival time only are useful in this approach, as they are sufficient both to provide candidate CRE events  and to determine the direction from which they are arriving.
}
\maketitle

\section{Introduction}
\label{intro}
The primary ultra-high energy cosmic-ray (UHECR) particles arriving at the Earth are very rare 
as their number falls steeply with increasing energy. 
Those which interact strongly or electromagnetically do not reach 
the surface of the Earth, but instead initiate a cascade of secondary particles in the atmosphere. 
These secondary particles reach the ground as a shower of relatively low energy muons, electrons and photons and are detectable in cosmic-ray
observatories. The most extended showers from a single cosmic-ray particle may, 
at the ground level, have a diameter of 3-4~km~\cite{uhecr_review}. The largest cosmic-ray observatories like Auger~\cite{auger} 
and Telescope Array~\cite{telescope_array} have their detectors spread over very larger areas
(3000~km$^{2}$ and 700~km$^{2}$, respectively) to ensure that such showers are fully contained.
This coverage is however not sufficient in case the shower starts not at the top of the Earth
atmosphere but at a significantly larger distance~\cite{SPS_Erber,SPS_McBreen}. Correlated bunches of particles may be initiated by 
UHECR photons and decays of hypothetical heavy and energetic Dark Matter particles. 

\begin{figure}[tb]
        \centering
        \includegraphics[width=8cm]{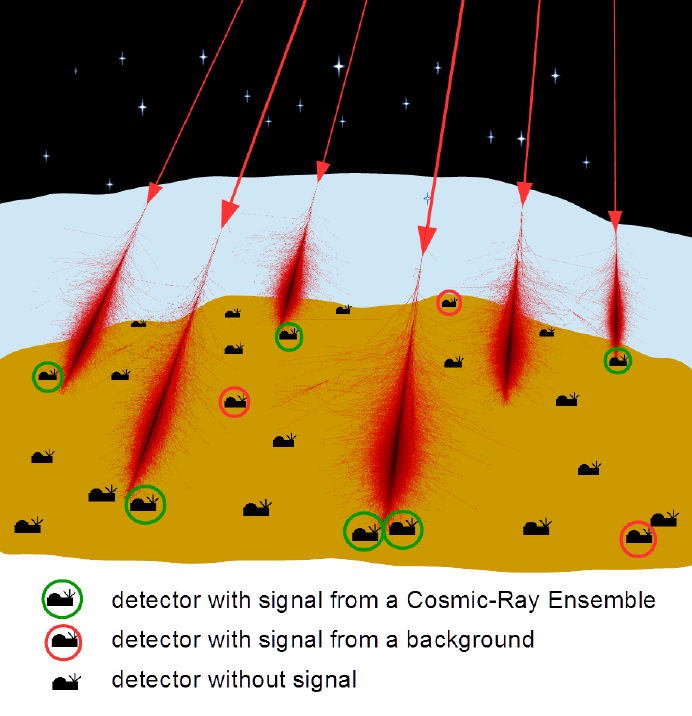}
        \caption{Detection of the Cosmic-Ray Ensemble (CRE) in distributed cosmic-ray detectors.}
        \label{fig_detection}
\end{figure}

Simulations of Super pre-showers (SPS) initiated by a photon and 
starting near the Sun show that in such a scenario the secondary photons near the Earth would form a disk with a diameter of hundreds 
of~km but with thickness of a few meters only~\cite{dithal_icrc2017}. 
In this case many photons arrive at almost the same time on top 
of the Earth atmosphere. Each of them has only a fraction of the energy 
of the primary photon, but still they may form numerous individual cascades.
The energy of these cascades is at least an order of magnitude lower 
than that of the primary particle and the distance between them may be
a few~km or more. Detection of such small showers, 
even if strongly correlated in time, 
by the traditional observatories like Auger may be impossible.
If in the pre-shower only a few particles are created it is 
most probable that in the area covered by the observatory 
only one shower appears, 
and it would not differ from showers generated by a single cosmic ray.
Several small showers may develop in the area covered by the observatory,
if in the pre-shower the number of particles is much larger. 
However, in this case
the energy of these particles has to be relatively small, thus the signals
in the detectors may have a pattern too different from a typical
high energy cascade to trigger recording of the event.


\begin{figure}[tb]
	\centering
	\includegraphics[width=8cm]{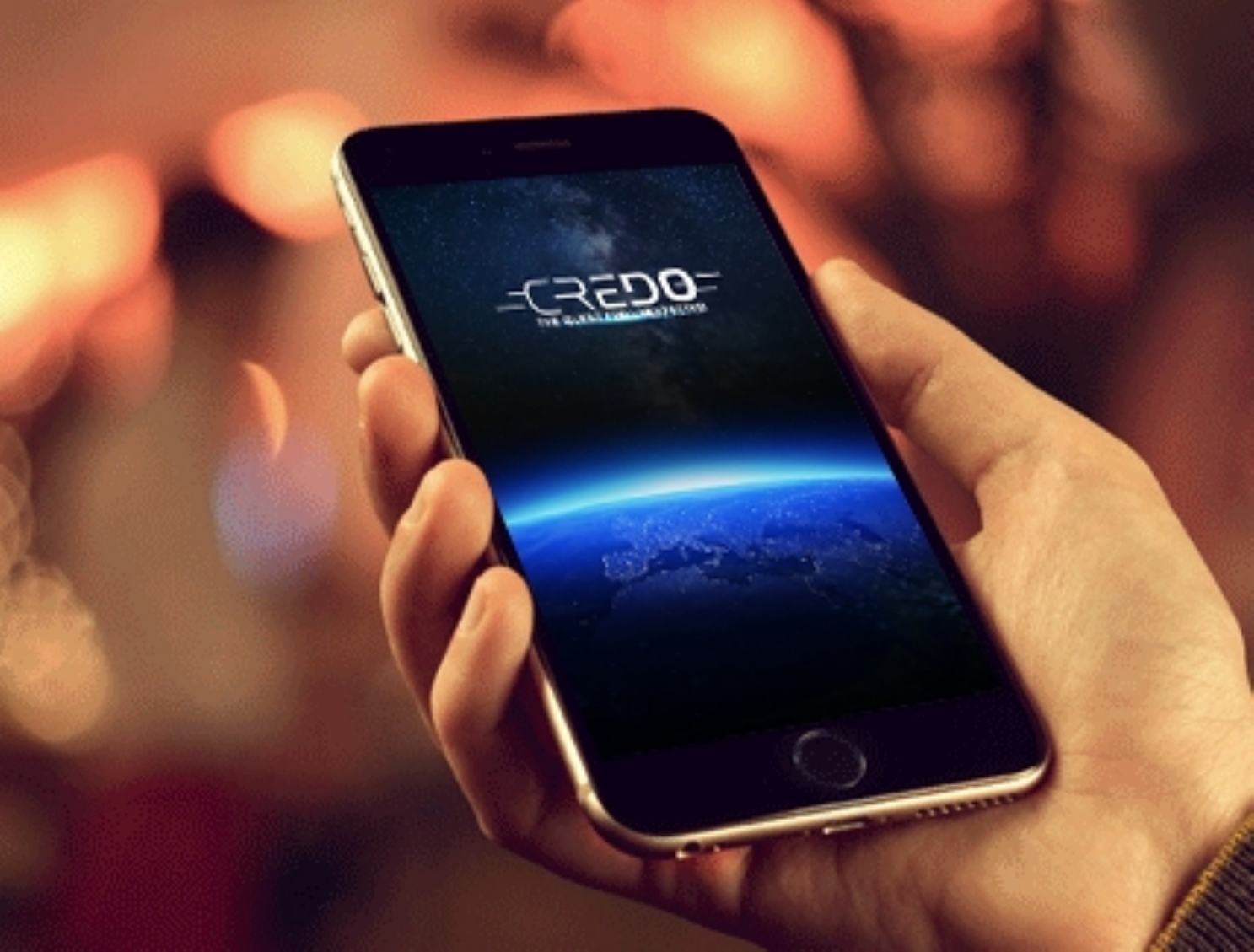}
	\caption{A smartphone with the CREDO application.}
	\label{fig_phone}      
\end{figure}

\begin{figure}[tb]
	\centering
	\includegraphics[width=7cm]{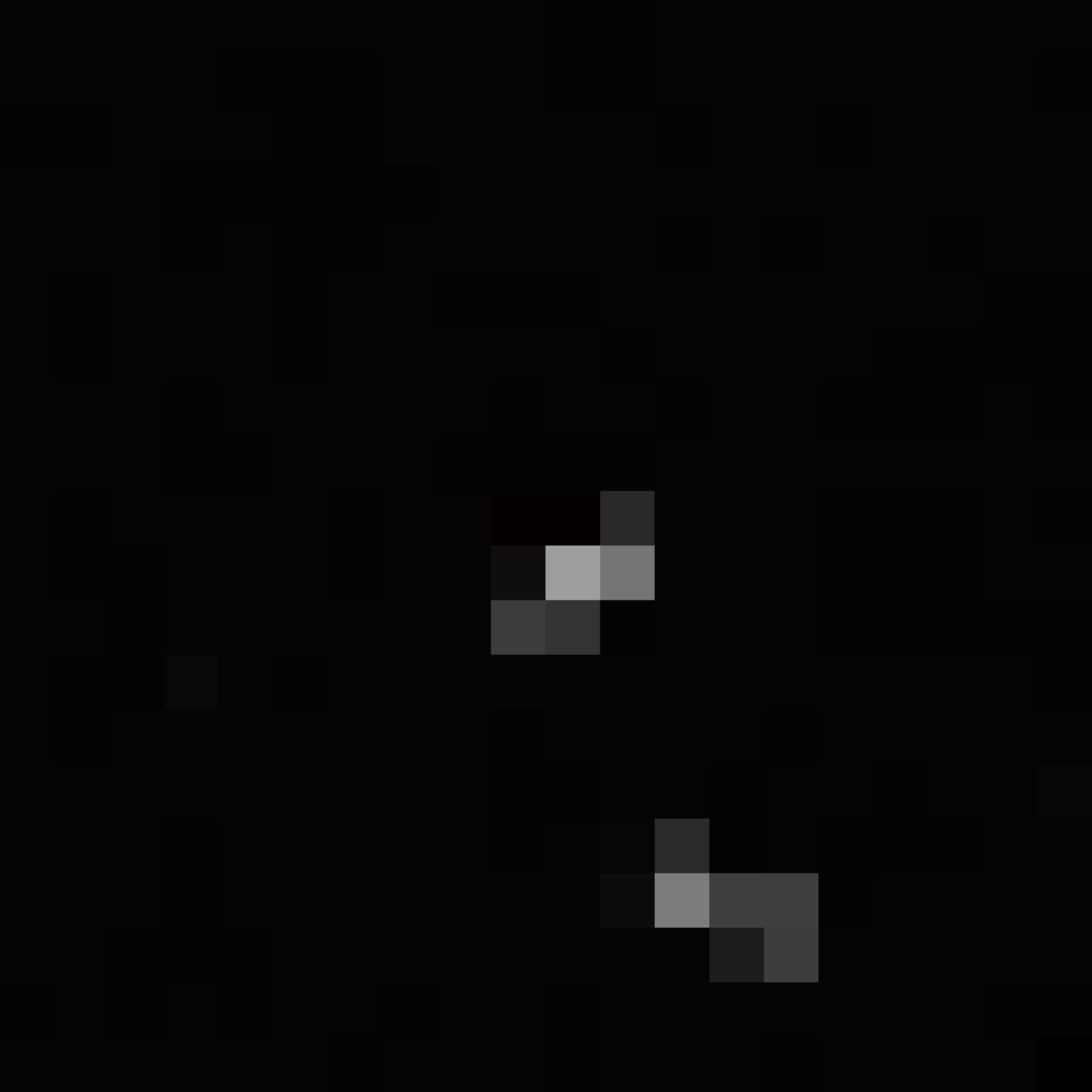}\\
	\vspace*{0.5cm}
        \includegraphics[width=7cm]{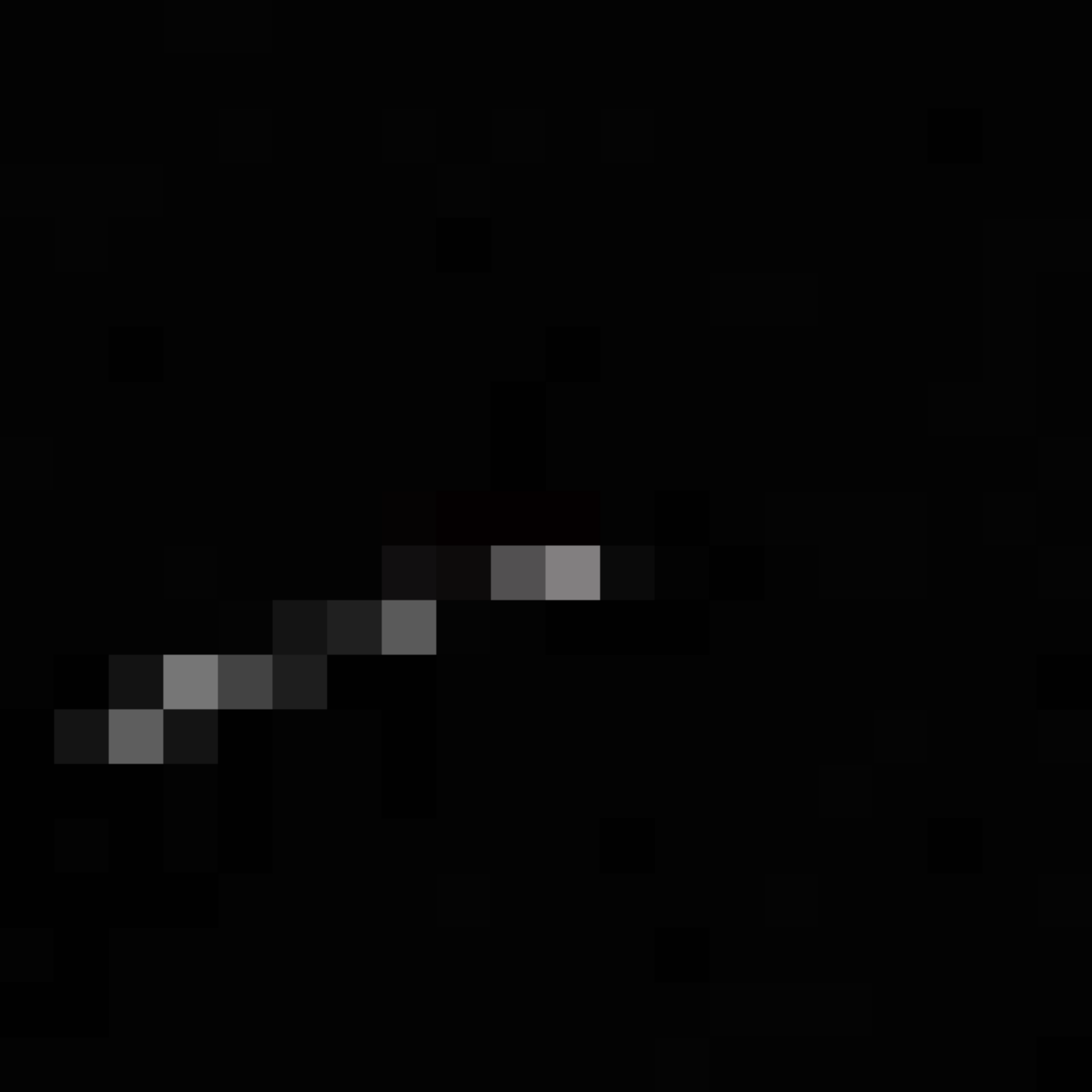}
	\caption{Examples of images taken with the completely covered camera of a smartphone using the CREDO application.}
	\label{fig_image}      
\end{figure}

\begin{figure}[tb]
	\centering
	\includegraphics[width=8cm]{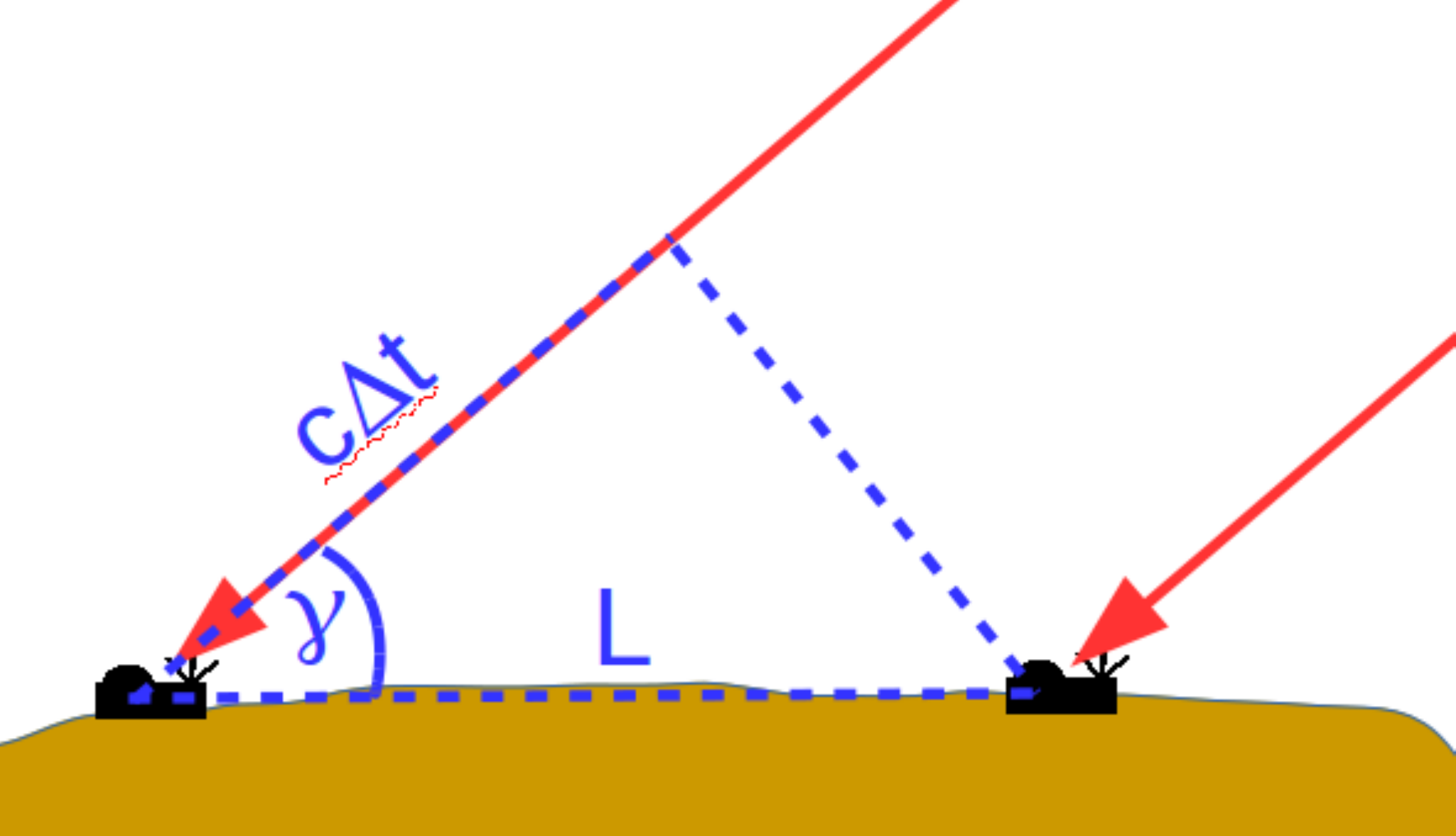}
	\caption{Estimation in two dimensions of the direction of an incoming Cosmic-Ray Ensemble on the basis of position and time registration in two detectors.}
	\label{fig_angle}      
\end{figure}

\section{Cosmic-Ray Extremely Distributed Observatory}
\label{credo}

Particles from a Cosmic-Ray Ensemble may arrive in places 
hundreds or even thousands~km distant and produce many individual 
cascades. Their signature would be the coincidence 
in time. Consequently, the detectors used to register them
need to cover the largest area possible, but may be quite sparse. 
It is not necessary to register many signals from individual cascades, 
it's better to detect signals from several different cascades,
even if it would be a single measurement for each cascade.
The best results can be achieved when the information from detectors 
spread over a country, continent or even the whole Earth are collected 
and analyzed together. The idea of CRE detection is presented 
in Figure~\ref{fig_detection}. The individual cascade leaves signal 
in one or a few detector stations. There are stations in which no signal
is registered in the analyzed time interval and also the stations 
with a background signal - for example cosmic rays not connected with CRE
or from decays of radioactive nuclei. 

The detectors registering the arrival time of 
particles are sufficient for testing the coincidences. 
However, rejection of backgrounds play an important role
in extraction of the signal and measurements of CRE properties
described in Section~\ref{triangulation}. Assuming that CRE initiates 
several small cascades it's reasonable to require a 
hardware based local coincidences of 
individual detectors within a station or a coincidence of nearby
stations. This way contribution from uncorrelated background 
signals (mainly radioactive decays) can be suppressed. 
It does not solve the problem of small cascades from cosmic 
rays not belonging to the Cosmic-Ray Ensemble - but their contribution,
at least partially, can be removed in the analysis.

The minimal hardware requirements are so easy to fulfill that 
the simplest of detectors capable of registering the time of a passage of 
a particle from a cosmic shower can be useful. The most popular
of such devices is a smartphone equipped with a camera and 
a GPS module. It can work as a cosmic-ray detector 
if the camera lens is covered
(to prevent external light) and an appropriate application is running.

The idea of using smartphones to detect cosmic rays is not new.
There are two groups CRAYFIS~\cite{crayfis_web,crayfis_arxiv} 
and DECO~\cite{deco_web, deco_icrc2015} which developed their custom
applications for this purpose for several years. However, their software
is proprietary thus the Open Source CREDO application 
(Figure~\ref{fig_phone}) was also created~\cite{credo_app}. 

The CREDO application works when the smartphone is attached 
to a power supply (to prevent discharging of the battery)
and the camera is protected from light. The pictures 
similar to those presented in Figure~\ref{fig_image} are constantly
taken and those with a pattern ressembling that from a cosmic ray are
sent to the server over the Internet. The information from 
the smartphones is stored and a summary of the activities 
is provided in a WWW page~\cite{credo_server}.
Currently, the work on prototypes of simple and inexpensive desktop type
cosmic-ray detectors is also ongoing.

\section{Time Based Triangulation}
\label{triangulation}

The Cosmic-Ray Ensemble created at a point very far from the Earth 
arrives as a front of almost parallel moving particles. 
In this case simple measurement of the position of the detector 
and the time when such particle arrives is sufficient 
to determine the direction of the front, as shown in Figure~\ref{fig_angle}.
Assuming the detected particles are traveling with the speed of light, $c$, 
the registration of the times  $t_{1}$ and $t_{2}$  in two detectors 
which are placed at the distance $L$, enables calculation of  
the inclination of the front, $\gamma$, as:
\begin{equation}
cos(\gamma) = c(t_{1}-t_{2}) / L.
\end{equation}
This angle calculated in  two dimensions  is in fact a half of the apex 
angle of the cone that is formed by all possible directions of the front.
Two detectors (stations) are not sufficient to determine
the direction of the front in 3D. However it is enough to add 
a third station with a signal, to construct two cones of possible
directions of the front, with a common apex. Then there are two 
lines - intersections of the cones - and that of them
which is pointing above the ground
is the correct one. This way even only three station with signals 
from CRE with a flat front after relatively simple calculations of the Time
Based Triangulation (TBT) method allow the estimation of the direction of the CRE.  
However, in the presence of ia background signal in even one of these stations  
the results of calculations are severely distorted. 
Frequently, the estimated cones may have no
intersection, either because a small cone is inside a bigger one or 
the two small cones are separated. In most cases, however, TBT calculations
give a direction differing greatly from the correct one. 
It is thus necessary to test all combinations of three stations and analyze 
calculated directions to find the CRE direction.

\begin{figure*}[tb]
        \centering
        \includegraphics[width=14cm]{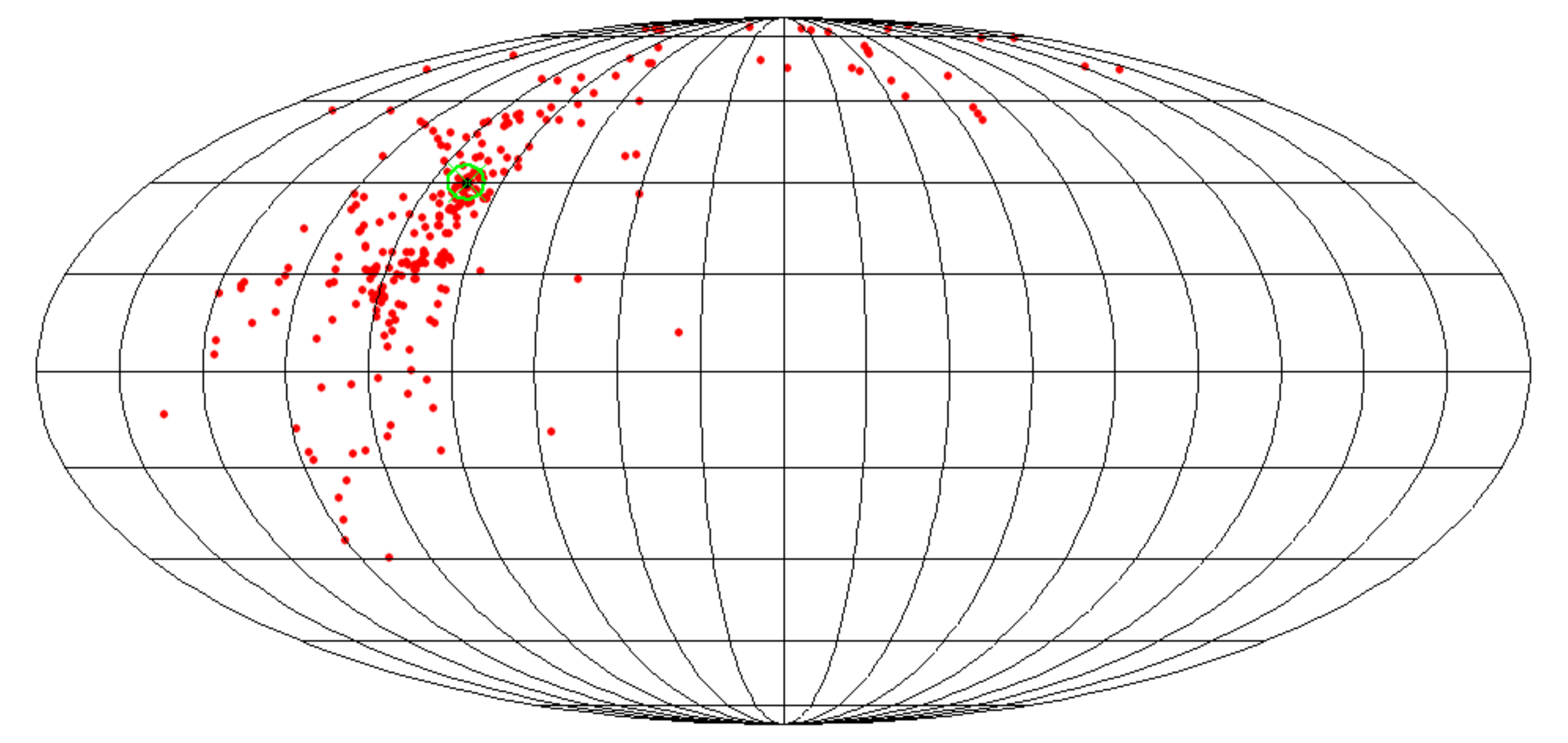}\\
        \vspace*{0.5cm}
        \includegraphics[width=11.2cm]{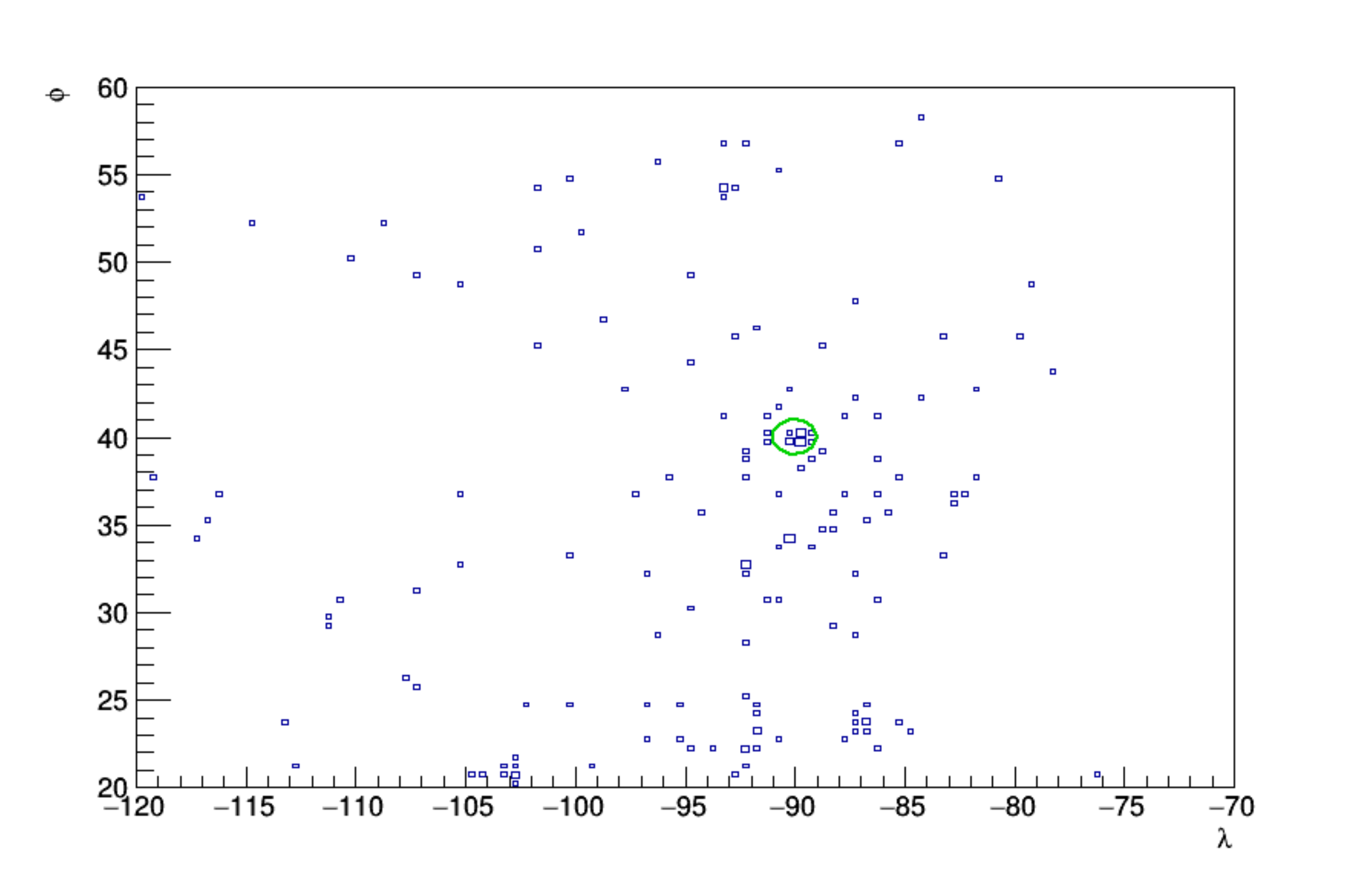}
        \caption{Estimation of possible directions of
an incoming Cosmic-Ray Ensemble using the Time Based Triangulation method.
Simulated signals from CRE were detected in 5 stations
while in another 10 stations randomly generated background signals
are present. In the upper plot all estimated directions 
in geographic coordinates system are shown
in Mollweide projection. The correct CRE direction is denoted
by the green circle, red points represent false estimates,
obtained from calculations using at least one background
signal. In the lower plot for selected range of  latitude, $\phi$,
and longitude, $\lambda$, the calculated directions are shown as
boxes, the correct ones are in the green circle.
}
        \label{fig_map}
\end{figure*}

\section{Simple CRE simulation}
\label{simulation}
The accuracy of calculations in the Time Based Triangulation method
was tested using simple simulations of a Super-preshower starting
very far from the Earth and thus approximated by a flat front of particles.
The positions of hypothetical detectors are randomly generated
in the area approximately representing the territory of USA.
The exact arrival time is randomly smeared with a width of
1/3000000~s, an uncertainty coresponding to the 1~km position
error along the SPS direcion. 
Signals from the SPS is simulated always in 5 stations, but 
in addition in some number of other stations background signals
are randomly generated in the time interval around the  arrival
time of the SPS. In Figure~\ref{fig_map} results of TBT reconstruction
for one event in which SPS signals and 
background are present in 5 and 10 stations, respectively, are shown.
The calculations were performed for all ~15!/(3!(15-3)!)~=~455
combinations of three stations with signals, but only in 5!/(3!(5-3)!)~=~10
combinations in all selected stations SPS signal is recorded.
Even if one station contains signal from a background the result of
TBT calculations can be completly wrong.
In the grid of geographical coordinates (Figure~\ref{fig_map} upper plot)
many calculated directions are thus false 
and are spread over almost a half of the solid angle.
In a closer examination of the angular area more focused 
on the real direction of SPS we can see an evident cluster 
of results near the correct direction and a less dense population 
around it. The calculated values from real signals
are quite precise, the errors do not exceed 1$^{\circ}$
thus for small and moderate background fractions the cluster
at the SPS directions should be easy to identify.

In the presented in Figure~\ref{fig_map} 
case of 5 real and 10 background signals 
the correct SPS direction can be guessed after 
checking the plot "by eye", but in more difficult cases 
much more elaborate data analysis is necessary.
In simulations with smaller number of stations with a background 
the density of false directions is lower and the correct 
direction is more pronounced. For larger number of stations 
with backgrounds the cluster showing the correct direction 
becomes less visible and may be difficult to find. It is thus 
very important to keep the ratio of stations 
with real signal to those with a background signal as high as possible.

\section{Summary}
\label{summary}

An obvious extension to the measurements of cosmic-ray showeres initiated
by a single high energy particle are studies of Cosmic-Ray Ensembles.
They may give an opportunity to discover new phenomena including
new types of interactions or decays of massive particles.
The purpose of CREDO is to collect all available cosmic-ray data 
and analyze correlations of signals in detectors spread over 
large distances. In this case detectors as simple as smartphones
may be useful, thus an Open-source CREDO application was created.
For measurements of electromagnetic Super-preshowers information
from even small tabletop detectors may be sufficient.
The Time Based Triangulation method allows calculation of 
direction of a SPS if only the background rate is not too large.

\section*{Acknowledgments}
This research has been supported in part by the Governments of Czechia, 
Hungary, Poland and Slovakia through Visegrad Grant 21720040 from International Visegrad Fund and by the PLGrid Infrastructure. 
We thank the staff at ACC Cyfronet AGH-UST for their helpful support.
CREDO application is developed in Cracow University of Technology.

\clearpage
\end{document}